\documentclass{PoS}

\title{$a_{0}(980)$ as a companion pole of $a_{0}(1450)$}

\ShortTitle{The $a_0(980)$ revisited}

\author{\speaker{Thomas Wolkanowski}\\
       Institute for Theoretical Physics, Goethe University, D-60438 Frankfurt am Main, Germany\\
       E-mail: \email{wolkanowski@th.physik.uni-frankfurt.de}}

\author{Francesco Giacosa\\
     Institute for Theoretical Physics, Goethe University, D-60438 Frankfurt am Main, Germany\\
     Institute of Physics, Jan Kochanowski University, PL-25406 Kielce, Poland\\
     E-mail: \email{fgiacosa@ujk.edu.pl}}

\abstract{Light scalar hadrons are often understood as dynamically generated resonances. These
arise as `companion poles' in the propagators of $q\bar{q}$ seed states 
when accounting for meson-loop contributions to the self-energies of the latter. 
Following this idea, we demonstrate that for the scalar--isovector state $a_{0}(1450)$ the full one-loop propagator 
has two poles: a pole of the seed state $a_{0}(1450)$ and a companion pole corresponding to $a_{0}(980)$. 
The positions of these poles are studied by varying the relative coupling strength between 
the non-derivative and derivative parts of the interactions.}

\FullConference{The 8th International Workshop on Chiral Dynamics \\
		 29 June 2015 - 03 July 2015\\
		 Pisa, Italy}

\begin{document}

\section{\label{sec:section1}Introduction}

Intense research during the past decades has demonstrated 
that the majority of mesons can be understood as being predominantly 
$q\bar{q}$ states \cite{olive}. However, various unconventional mesonic states 
such as glueballs, hybrids, and four-quark states are expected \cite{amslerrev}. 
Along this line, a concept of `dynamically generated' states was put forward in Refs.\ 
\cite{dullemond,morgan,tornqvist,pennington}. Allthough there is not a generally accepted definition 
of dynamical generation \cite{giacosaDynamical}, other versions of this idea 
can also be found in Refs. \cite{tornclose,pelaez,oller,2006beveren}. 

The general idea is summarized in the following: Consider, for instance, 
a single seed state, \emph{e.g.}\ a $q\bar{q}$ meson with certain quantum numbers. 
This state interacts with other mesons, giving rise to loop contributions
in the corresponding self-energy and dressing its own full propagator. These contributions shift the corresponding pole 
of the seed state which moves away from the real axis and follows a certain trajectory 
in the appropriate unphysical Riemann sheet \cite{peierls}. Moreover, new poles may appear. 
The latter are sometimes denoted as companion poles. If one of them happens to be situated sufficiently 
close to the physical region, $i.e.$, the real axis, it could correspond to a dynamically generated resonance. 
As a consequence, we are left with two resonances emerging from a single seed state. 
In this work, we aim to discuss this mechanism in the context of the resonance $a_{0}(980)$.

From what was found for example in Refs.\ \cite{close,tqmix,eLSM1,eLSM2} 
the scalar resonances $f_{0}(1370)$, $f_{0}(1500)$, $K_{0}^{\ast}(1430)$, and $a_{0}(1450)$
seem to be predominantly ordinary $q\bar{q}$ states. On the other hand, the light scalar states 
$f_{0}(500)$, $f_{0}(980)$, $K_{0}^{\ast}(800)$, and $a_{0}(980)$ are (most likely) predominantly 
something different (see \emph{e.g.} Refs.\ \cite{oller,tqmix,jaffe,maiani,hooft,fariborz,rodriguez,molecular} 
and refs. therein). In fact, we have proven in a recent publication \cite{thomasPRD} that for the heavy 
scalar--isovector seed state $a_{0}(1450)$, the coupling of this state to $\pi\eta$, $K\bar{K}$, and 
$\pi\eta^{\prime}$ is capable to dynamically generate the light state $a_{0}(980)$. We will shortly outline 
the hadronic model that was applied there, which includes meson interactions via derivative 
and non-derivative couplings. We will then study the dependence of the pole structure with respect to the 
the relative coupling strength between the non-derivative and derivative parts of the interactions.

These proceedings are based on Ref. \cite{thomasPRD}. 
Our units are $\hbar=c=1$. The metric tensor is $\eta_{\mu \nu} = diag(+,-,-,-)$.

\section{\label{sec:section2}Formalism}

Following the earlier work of Ref. \cite{dullemond}, some seminal studies investigated the scalar sector 
in a unitarized quark model by including meson-loop contributions \cite{tornqvist,pennington}. They showed
that meson-loop effects may serve to explain the existence of the light scalar mesons. As can be seen 
from our discussion in Ref. \cite{thomasPRD}, the situation is however somewhat inconclusive 
regarding the number and the location of propagator poles and how to assign them to physical states. 
After reviewing the general approach, we present our way to model the scalar--isovector sector.

The main goal is the determination of the inverse propagator of a 
resonance after applying a Dyson resummation of loop contributions to the self-energy:
\begin{equation}
\Delta^{-1}(s)=s-m_{0}^{2}-\Pi(s)\ .
\end{equation}
Here, $m_{0}$ is the bare mass of the seed state and $\Pi(s)=\sum_{i}\Pi_{i}(s)$ is the self-energy.
The sum runs over the loops emerging from the coupling of the resonance to various mesons. 
The real part of $\Pi(s)$ on the real axis is related to the imaginary part by the 
dispersion relation
\begin{equation}
Re\Pi(s)=\frac{1}{\pi} -\hspace{-0.395cm}\int \mbox{d}s^{\prime} \ 
\frac{-Im\Pi(s^{\prime})}{s-s^{\prime}} \ .
\label{eq:disp}
\end{equation}
The actual modeling occurs in the particular expression of the imaginary part of $\Pi_i(s)$. 
According to the optical theorem, it corresponds to the partial decay width of the resonance into mesons 
in channel $i$, see Sec. \ref{sec:section3}. Furthermore, a form factor is usually introduced,
\begin{equation}
F_{i}(s)=\exp[-k_{i}^{2}(s)/\Lambda^{2}] \ ,
\end{equation}
where $\Lambda$ is a cutoff parameter and $k_{i}(s)$ is the absolute value of the 
three-momentum of the decay particles in the rest frame of the resonance:
\begin{equation}
k_{i}(s) = \frac{1}{2\sqrt{s}}\sqrt{s^{2}+(m_{i1}^{2}-m_{i2}^{2})^{2}-2(m_{i1}^{2}+m_{i2}^{2})s} \ .
\end{equation}
Here, $m_{i1},m_{i2}$ are the masses of the decay products, \emph{i.e.}, in our case
the pseudoscalar mesons. The function $F_{i}(s)$ guarantees that the imaginary part of $\Pi(s)$ vanishes 
sufficiently fast for $s \rightarrow \infty$.

The self-energy on the unphysical sheet(s), $\Pi^{c}(s)$, is obtained by analytic continuation:
\begin{eqnarray}
Disc\Pi(s) & = & 2i\lim_{\epsilon\rightarrow0^{+}}\sum_{i}
Im\Pi_{i}(s+i\epsilon) \ , \ \ \ s \in \mathbb{R} \ , \\
\Pi^{c}(s) & = & \Pi(s)+Disc\Pi(s) \ .
\end{eqnarray}
In this work only the three sheets nearest to the physical region will be regarded (in the standard notation $\pi\eta\leftrightarrow$II, $K\bar{K}\leftrightarrow$III, $\pi\eta^{\prime}\leftrightarrow$VI).

\section{A simple effective model with derivative interactions}
\label{sec:section3}

\subsection{Lessons from previous works}
\label{sec:sec3.1}

A first attempt to incorporate the mentioned mechanism of dynamical generation for the scalar states with $I=1$
in a consistent scheme \cite{eLSM1,eLSM2,stani} was presented in Refs.\ \cite{procEEF70,proceqcd}. 
There, the seed state was assigned to be (in the mass region of) the $a_{0}(1450)$, 
and the $s$-dependence of the amplitudes was completely neglected (apart from the cutoff dependence), 
yielding a width of the seed state which is too small. Moreover, \emph{no} additional pole for the $a_{0}(980)$ was 
dynamically generated.

It was then demonstrated in Ref. \cite{thomasPRD} that it is however possible 
to obtain a narrow resonance with mass around $1$ GeV, the pole coordinates of which 
fit quite well with those of the physical $a_{0}(980)$ resonance, 
and \emph{simultaneously} obtain a pole for the seed state in agreement with that for the $a_{0}(1450)$ \cite{olive}. 
An important requirement seems to be the inclusion of $s$-dependent amplitudes and derivative interaction terms in the 
Lagrangian, respectively.

\subsection{Effective model with both non-derivative and derivative interactions}

We now consider the effective model for the isovector states from Ref. \cite{thomasPRD} which contains non-derivative and derivative interactions. 
The Lagrangian is given by the sum of the following terms:
\begin{eqnarray}
\mathcal{L}_{a_{0}\eta\pi} & = & A_{1}a_{0}^{0}\eta\pi^{0}
+B_{1}a_{0}^{0}\partial_{\mu}\eta\partial^{\mu}\pi^{0} \ , \label{eq:Lag_eff} \\
\mathcal{L}_{a_{0}\eta^{\prime}\pi} & = & A_{2}a_{0}^{0}\eta^{\prime}\pi^{0}
+B_{2}a_{0}^{0}\partial_{\mu}\eta^{\prime}\partial^{\mu}\pi^{0} \ , \nonumber \\
\mathcal{L}_{a_{0}K\bar{K}} & = & A_{3}a_{0}^{0}(K^{0}\bar{K}^{0}-K^{-}K^{+})
+B_{3}a_{0}^{0}(\partial_{\mu}K^{0}\partial^{\mu}\bar{K}^{0}-\partial_{\mu}K^{-}\partial^{\mu}K^{+}) \nonumber \ .
\end{eqnarray}
Then, Eq.\ (\ref{eq:Lag_eff})
gives rise to the following $s$-dependent amplitudes:
\begin{equation}
\mathcal{M}_{i}^{eff}(s) = \left[A_{i}-\frac{1}{2} B_{i}\left(s-m_{i1}^{2}-m_{i2}^{2}\right)
\right] F_{i}(s) \ ,
\end{equation}
where we have already included a regularization function $F_{i}(s)$ as defined in Sec.\
\ref{sec:section2}.

The imaginary part of the one-loop self-energy\footnote{The one-loop approximation for the self-energy is
quite reliable, since vertex corrections can be shown to have a negligible effect \cite{jonas}.} is computed by using the optical theorem,
\begin{equation}
Im\Pi_{i}(s) = - \sqrt{s}\, \Gamma_{i}^{tree}(s) = - \frac{k_{i}(s)}{8\pi 
\sqrt{s}}|\mbox{--}i\mathcal{M}^{eff}_{i}(s)|^{2}\Theta(s-s_{th,i}) \ ,
\label{eq:optical}
\end{equation}
and the real part comes from the dispersion relation in Eq. (\ref{eq:disp}). The step function 
ensures that the decay channel $i$ contributes only when the squared energy of the
resonance exceeds the threshold value $s_{th,i}$. Notice that from a careful analysis
we showed the necessity to introduce subtractions that are not visible here -- for a detailed presentation of this 
issue see Ref. \cite{thomasPRD}.

Our fitting procedure was aimed to find a set of parameters $\{m_{0},\ \Lambda\}$ for which $(i)$ 
two poles appropriate for the $I=1$ resonances $a_0(980)$ and $a_0(1450)$ can be found,
\emph{i.e.}, poles that lie on the second and sixth sheet, respectively, and $(ii)$ the six coupling constants 
$A_i$, $B_i$ $(i=1,2,3)$ produce branching ratios of $a_{0}(1450)$ in good agreement with the central values of the
experimental branching ratios \cite{olive}. We obtained \cite{thomasPRD}:
\begin{eqnarray}
m_{0}=1.15 \ \mbox{GeV} \ , \ \ \ \ \ \ \ & & \ \Lambda = 0.6 \ \mbox{GeV} \ , \\
A_{1} = 2.52 \ \mbox{GeV} \ , \ \ \ \ \ \ \ & & B_{1} = -8.07 \ \mbox{GeV}^{-1} \ , \label{eq:AiBi} \\
A_{2} = 9.27 \ \mbox{GeV} \ , \ \ \ \ \ \ \ & & B_{2} = 9.25 \ \mbox{GeV}^{-1} \ , 
\nonumber \\
A_{3} = -6.56 \ \mbox{GeV} \ , \ \ \ \ & & B_{3} = -1.54 \ \mbox{GeV}^{-1} \ . \nonumber
\end{eqnarray}
By using the tree-level decay widths obtained from the optical theorem (\ref{eq:optical}) 
at the peak value of the spectral function above $1$ GeV, $m_{a_{0}(1450)}^{peak}=1.419$ GeV:
\begin{equation}
\frac{\Gamma_{a_{0}(1450)\rightarrow\eta^{\prime}\pi}^{tree}}{
\Gamma_{a_{0}(1450)\rightarrow\eta\pi}^{tree}}\simeq0.44 \ , \ \ \ 
\frac{\Gamma_{a_{0}(1450)\rightarrow K\bar{K}}^{tree}}{
\Gamma_{a_{0}(1450)\rightarrow\eta\pi}^{tree}}\simeq0.96 \ .
\end{equation}
The pole corresponding to the $a_{0}(980)$ has coordinates\footnote{We apply the usual parameterization for propagator poles, 
$s_{pole}=m_{pole}^{2}-i\hspace{0.02cm}m_{pole}\Gamma_{pole}$ \ .}
\begin{equation}
\sqrt{s_{pole}}\big|_{a_{0}(980)} = (0.970-i\hspace{0.02cm}0.045) \ \mbox{GeV} \ ,
\end{equation}
\emph{i.e.}, we find the $a_{0}(980)$ to have a mass of $m_{pole}^{a_{0}(980)}=0.969$ GeV 
and a width of $\Gamma_{pole}^{a_{0}(980)}=0.090$ GeV. For the resonance above $1$ GeV we get
\begin{equation}
\sqrt{s_{pole}}\big|_{a_{0}(1450)} = (1.456-i\hspace{0.02cm}0.134)  \ \mbox{GeV} \ ,
\end{equation}
or $m_{pole}^{a_{0}(1450)}=1.450$ GeV and $\Gamma_{pole}^{a_{0}(1450)}=0.270$ GeV. 

\section{Results}
\label{sec:section4}

\begin{figure}
\hspace{2.2cm}
\includegraphics[scale=1.0]{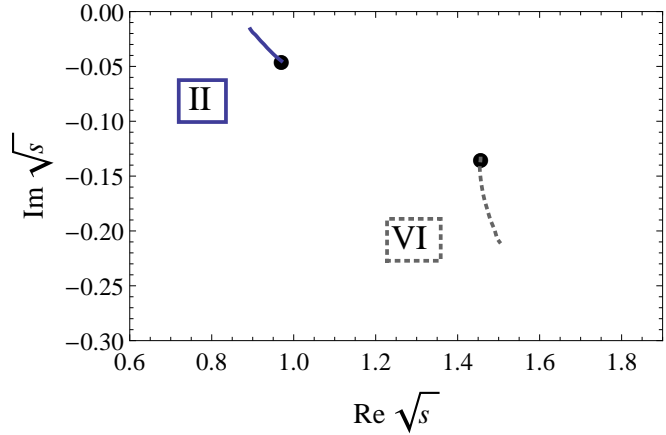}
\caption{Pole structure of our effective model in dependence of $\delta$. 
Black dots indicate the position of the poles for $\delta=1.0$. 
The roman number indicates on which sheet the respective pole can be found.}
\label{fig:delta}
\end{figure}

In the following, we do something different with respect to Ref. \cite{thomasPRD}. We introduce a dimensionless 
parameter $\delta \in [0,1]$ and replace the derivative coupling constants in 
Eq.\ (\ref{eq:Lag_eff}) by $B_{i}^{2}\rightarrow \delta B_{i}^{2}$. In consequence, for $\delta =0$ the self-energy 
contains only non-derivative interactions, while for $\delta=1.0$ we reproduce the poles stated before. 
Increasing $\delta$ from zero to one, the derivative interaction is successively increased and we can monitor 
in a controlled manner how the pole structure changes. The result can be seen in Fig. \ref{fig:delta}.

It turns out that it is 
possible to obtain two poles even for vanishing $\delta$ where the derivative interactions 
give no contribution. In this case the real part of the corresponding pole for $a_{0}(980)$ (second sheet) is maybe too small, 
but the imaginary part is definitely too small. On the other hand, the latter of the pole for $a_{0}(1450)$ (sixth sheet) is obviously 
too large. Driving $\delta\rightarrow1.0$, both poles reach their final positions in different ways: The pole 
on the second sheet gains changes in both its real and imaginary parts; both are increased. Concerning the pole on the sixth 
sheet, the real and imaginary parts decrease, but the latter is more affected by a change of $\delta$. For the spectral function see Ref. \cite{thomasPRD}

\section{Conclusions}
\label{sec:conc}

Our results demonstrate that it is in fact possible to correctly 
describe the resonances $a_{0}(980)$ and $a_{0}(1450)$ in a unique framework, 
where originally only a single $q\bar{q}$ seed state is present. The starting point is an effective Lagrangian 
that includes both derivative and non-derivative interaction terms, see Eq.\ (\ref{eq:Lag_eff}). The form of the Lagrangian 
is inspired by the one of the extended Linear Sigma Model (eLSM). From our variation 
of the overall coupling strength $\delta$ we furthermore showed that both terms seem to be equally important.

The presented mechanism of dynamical generation (of light scalar mesons) may be extended in 
two directions: $(i)$ One could study the isodoublet, \emph{i.e.}, by describing the resonances 
$K_{0}^{\ast}(800)$ and $K_{0}^{\ast}(1430)$ in a similar unified framework (this we already started, see Ref. \cite{ourkappa}). 
The pole of $K_{0}^{\ast}(800)$ is not yet very well known and there is need of 
improved analyses. $(ii)$ Furthermore, the scalar--isoscalar sector could be investigated, 
where the resonances $f_{0}(500)$ and $f_{0}(980)$ should be dynamically generated, 
while $f_{0}(1370),$ $f_{0}(1500)$, and $f_{0}(1710)$ would be predominantly 
a non-strange quarkonium, a strange quarkonium, and a scalar glueball, respectively.

Another interesting project is the study of dynamical generation in the framework of 
resonances in the charmonium sector \cite{PenningtonCharm}, see for example Ref.\ \cite{brambilla} 
and refs.\ therein. Namely, a whole class of mesons, called $X$, $Y$, and $Z$ states, 
has been experimentally discovered but is so far not fully understood \cite{braaten,maianix}.
As shown in Ref.\ \cite{coitox} for the case of $X(3872)$, some of the $X$ and $Y$ states 
could emerge as companion poles of $q\bar{q}$ states.

\section*{Acknowledgements}
The authors thank Dirk H. Rischke, M.\ Pennington, J.\ Wambach, G.\ Pagliara, J.\ Reinhardt, D. D.\ Dietrich, 
R.\ Kami\'{n}ski, J. R.\ Pel\'{a}ez, and H.\ van Hees for useful discussions. T. W.\ 
acknowledges financial support from HGS-HIRe, F\&E GSI/GU, and HIC for FAIR Frankfurt.

\end{document}